# Stochastic modeling of Congress

M.V. Simkin and V.P. Roychowdhury
*Department of Electrical Engineering, University of California, Los Angeles, CA 90095-1594*

We analyze the dynamics of growth of the number of congressmen supporting the resolution HR1207 to audit the Federal Reserve. The plot of the total number of co-sponsors as a function of time is of "Devil's staircase" type. The distribution of the numbers of new co-sponsors joining during a particular day (step height) follows a power law. The distribution of the length of intervals between additions of new co-sponsors (step length) also follows a power law. We use a modification of Bak-Tang-Wiesenfeld sandpile model to simulate the dynamics of Congress and obtain a good agreement with the data.

The webpage [1] contains the names of all congressmen, who have co-sponsored the HR1207 resolution along with the date they co-signed. The plot of the number of co-sponsors as a function of date created using this data is shown in Figure 1(a). It is highly irregular with long periods without addition of any new co-sponsors interrupted by jumps, when many new co-sponsors join during a single day. Such a curve is known in mathematics as the "Devil's staircase" [2]. The staircase can be characterized by the distributions of step lengths and step heights. Figures 2 and 3 show such distributions for the staircase of Figure 1(a). The plots look linear in log-log coordinates, which suggest that the distributions follow a power law.

The power law distribution of the numbers of new co-sponsors joining during a day calls into mind the Bak-Tang-Wiesenfeld (BTW) sandpile model [3] where a similar power-law distribution of avalanches was observed. One may hope that some modification of the model can explain the dynamics of Congress. Our model is as follows. There is a *network of influence* in Congress through which representatives exert upon each other political pressure. When political pressure on certain congressman reaches a threshold, he co-sponsors the resolution. When a congressman co-sponsors the resolution, it exerts political pressure on other congressmen, whom he influences. In addition, there are calls on congressmen from their constituency which are analogous to randomly dropping grains of sand in the BTW model. The major difference with the BTW model is that a congressman cannot "topple" more than once regarding the resolution in question because after he sponsors it, he cannot do it again.

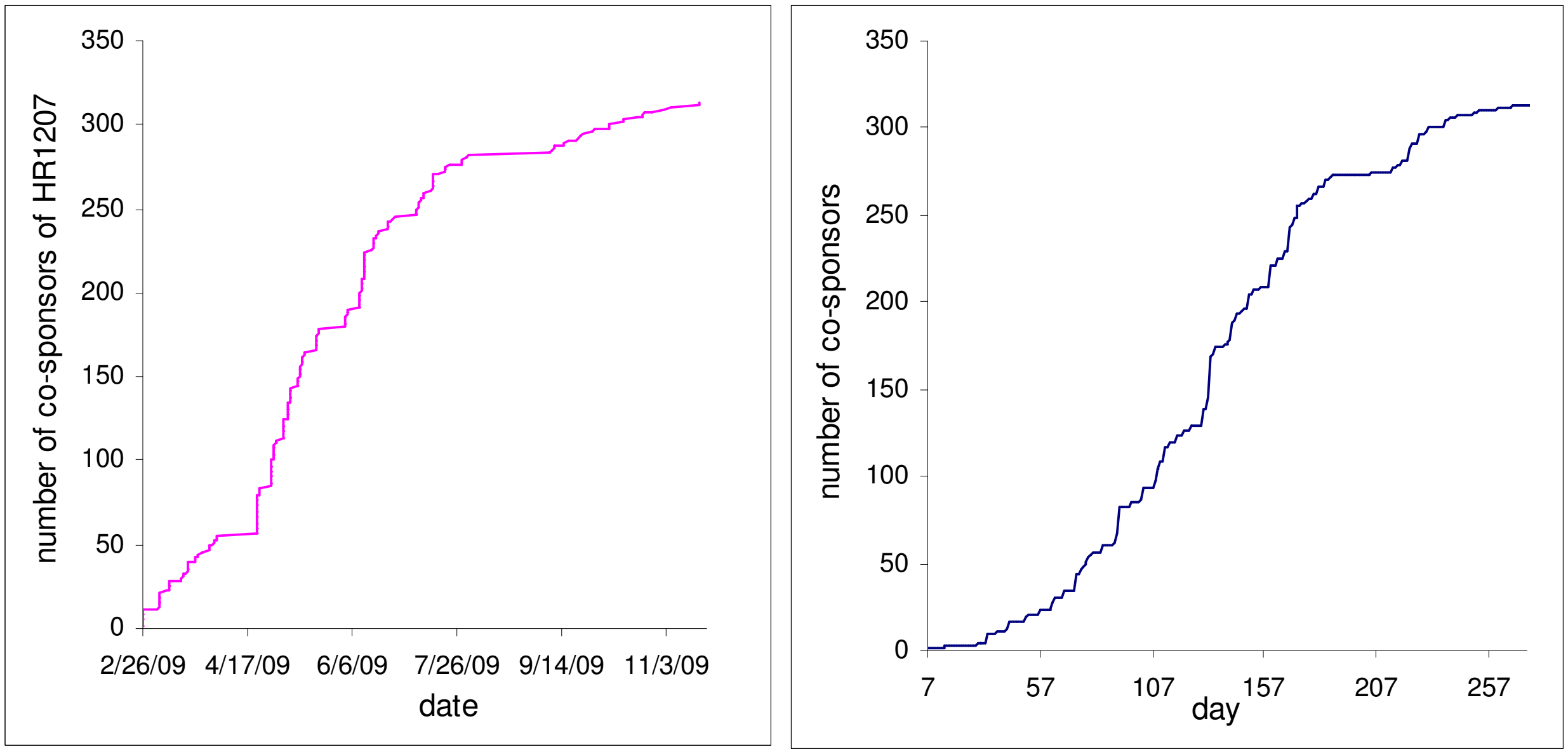

(a) (b)

**Figure 1**. (a) Devil's staircase represents the number of co-sponsors of Ron Paul's HR1207 resolution to audit the Federal Reserve as a function of date. As of November 19 2009, 313 of 435 congressmen co-sponsored the resolution. (b) Results of one of the simulations of congressional sandpile model for the same number of days.

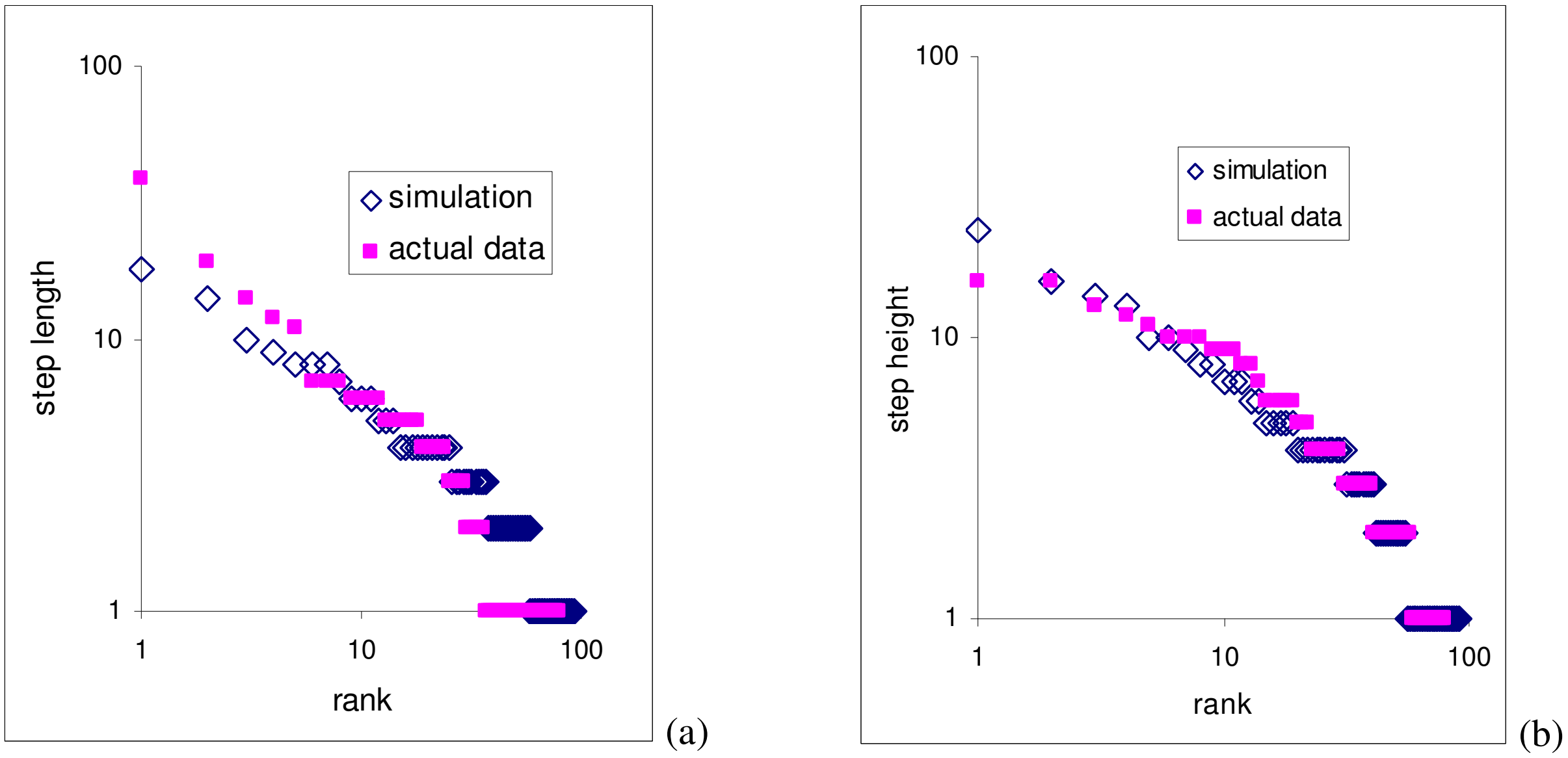

**Figure 2.** Rank-frequency distribution of: (a) the intervals between additions of new co-sponsors (step length); (b) the numbers of new co-sponsors joining during a day (step height)

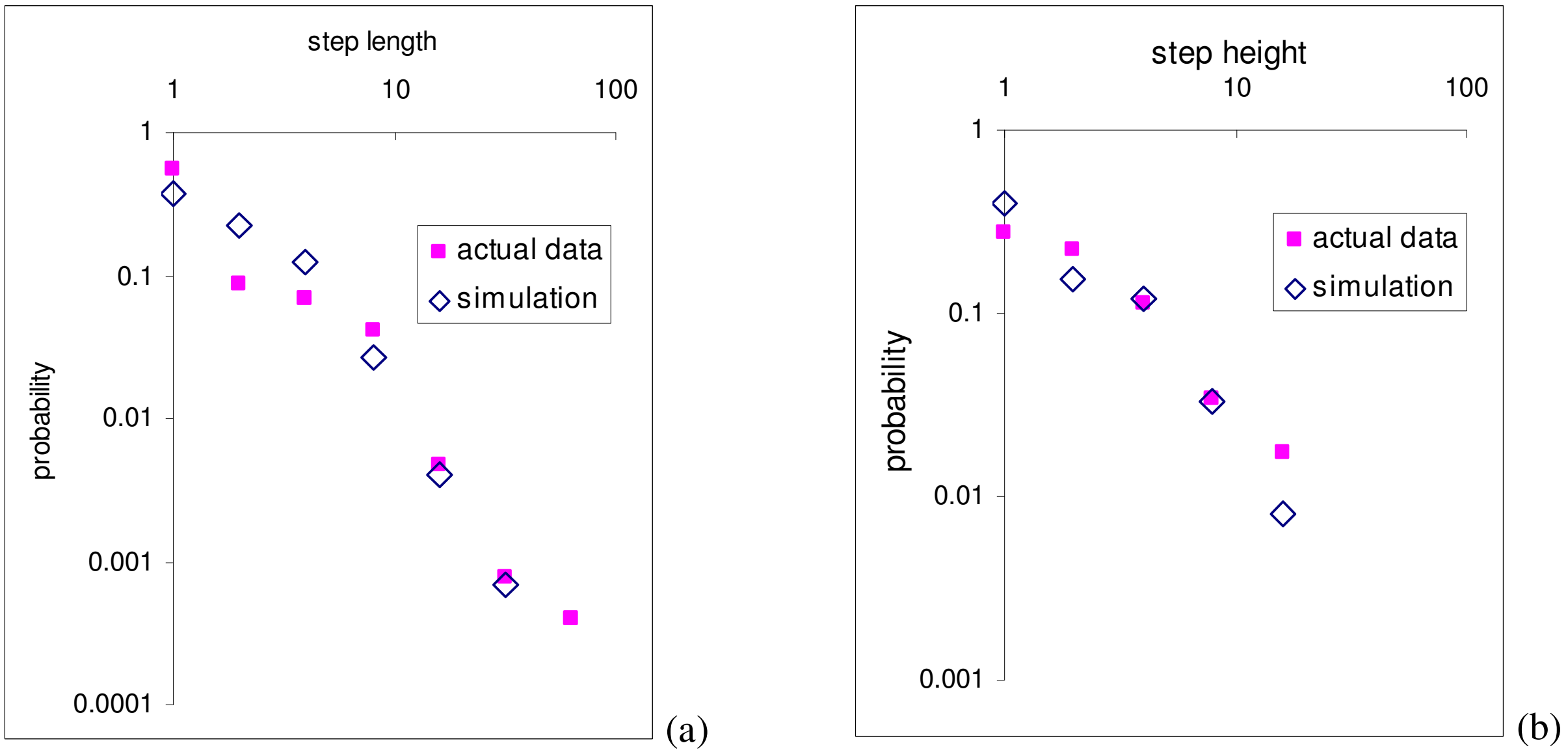

**Figure 3.** Probability density distribution of: (a) the intervals between additions of new co-sponsors (step length); (b) the numbers of new co-sponsors joining during a day (step height)

In our simulations we used a network where every one of 435 congressmen was influencing $k$ randomly selected congressmen and was influenced by $k$ randomly selected congressmen[1]. The "toppling" threshold for each congressman was set to $k$ units of political pressure. The calls from the constituency were Poisson distributed with the mean number of daily units of political pressure equal to $\lambda$. Thus in the process of our simulation for every day a Poisson-distributed random number $j$ is generated. Afterward the following procedure is repeated $j$ times. One congressman is selected at random out of 435 congressmen to receive a unit of political pressure from a constituent. In the case that he reaches the toppling threshold, he co-sponsors the resolution and transfers $k$ units of political pressure to $k$ congressmen, which he influences. One or more of those congressmen can reach the threshold, co-sponsor the resolution, and transfer political pressure to those they influence. If a congressman, who had already toppled, receives new units of political pressure, they just stay with him and have no further effect. We proceed to topple the congressmen until Congress is in a stationary state. At the end of the day, we record the number of new co-sponsors and proceed to the next day.

[1] We use this simplistic model because of a lack of available data on the influence network among the members of Congress. We could investigate a more realistic influence network process given such data.

We tried different values of $k$ and could obtain a good agreement with the data using $k = 2$ and $k = 3$. Figure 1(b) shows the staircase produced by one of such simulations $k = 2$ and $\lambda = 1.7$ (the $\lambda$ parameter merely rescales the time and thus can be tuned to have the required number of congressmen topple during the given number of days) . The plot starts with the day number 7 when we got the first toppling and continues for the same number of days (267) as the HR1207 co-sponsorship data that we have. Figures 2 and 3 show the step length and step height distributions and a good agreement between simulations and the actual data is evident.

We can get some insight into the behavior of our model using analytical methods. We used a random network of influence in our model. For such networks, a mean field theory is exact in the limit of infinite network size. The state of Congress for the case $k = 2$ is described by just two parameters: the fraction of untoppled congressmen with 0 units ($p_0$) and 1 unit ($p_1$). The fraction of toppled congressmen is determined by these two variables:

$$p_t = 1 - p_0 - p_1. \qquad (1)$$

Consider a congressional sandpile of large size $L$ to which $M$ units of political pressure had been added. Let us find $p_0$ and $p_1$ as functions of the average number of units per congressman $m = M/L$. We add a new unit of political pressure. If it falls on a toppled congressman – it has no further effect. If it falls on an untoppled 0-unit congressman (this happens with probability $p_0$) then

$$dp_0 = -dm;\ dp_1 = dm, \qquad (2)$$

where $dm = 1/L$. If the unit of pressure falls on an untoppled 1-unit congressman (this happens with probability $p_1$) this congressman topples. He transfers two units of pressure to two other congressmen, each of which is in a condition to topple (is untoppled 1-unit) with probability $p_1$. This starts a branching process (see Ref. [4]), and the average number of toppling in the second round is $2p_1$. The average number of topplings $\langle n \rangle$ in the ensuing avalanche is the sum of the numbers in a geometric progression with the ratio $r = 2p_1$: $\langle n \rangle = 1/(1-2p_1)$. An average avalanche displaces $2\langle n \rangle$ units. Obviously, $2\langle n \rangle p_0$ of them go to 0-unit congressmen changing them to the 1-unit state. Thus:

$$dp_0 = -2\langle n \rangle p_0 dm = -\frac{2p_0}{1-2p_1} dm. \qquad (3a)$$

Correspondingly $p_1$ increases by the same quantity. In addition, an avalanche topples $\langle n \rangle$ 1-unit congressmen. Thus:

$$dp_1 = \frac{2p_0}{1-2p_1} dm - \frac{1}{1-2p_1} dm = \frac{2p_0 - 1}{1-2p_1} dm. \qquad (3b)$$

Now we multiply the right hand sides of Equations (2) and (3) by the probability of each event, sum them and get:

$$\frac{dp_0}{dm} = -p_0 - p_1 \frac{2p_0}{1-2p_1} = -\frac{p_0}{1-2p_1}\ ; \qquad (4a)$$

$$\frac{dp_1}{dm} = p_0 + p_1 \frac{2p_0 - 1}{1-2p_1} = \frac{p_0 - p_1}{1-2p_1}. \qquad (4b)$$

The "analytical" curve in Figure 4(a) was computed using Eq. (1) and numerical solution of Eq. (4) with the initial conditions $p_0 = 1$ and $p_1 = 0$. The "simulation" curve in Figure 4(a) is the same simulation as the one shown in Figure 1(b), except the *x*-axis is rescaled by the factor $\lambda / L = 1.7/435$ and the *y*-axis by the factor $1/L = 1/435$.

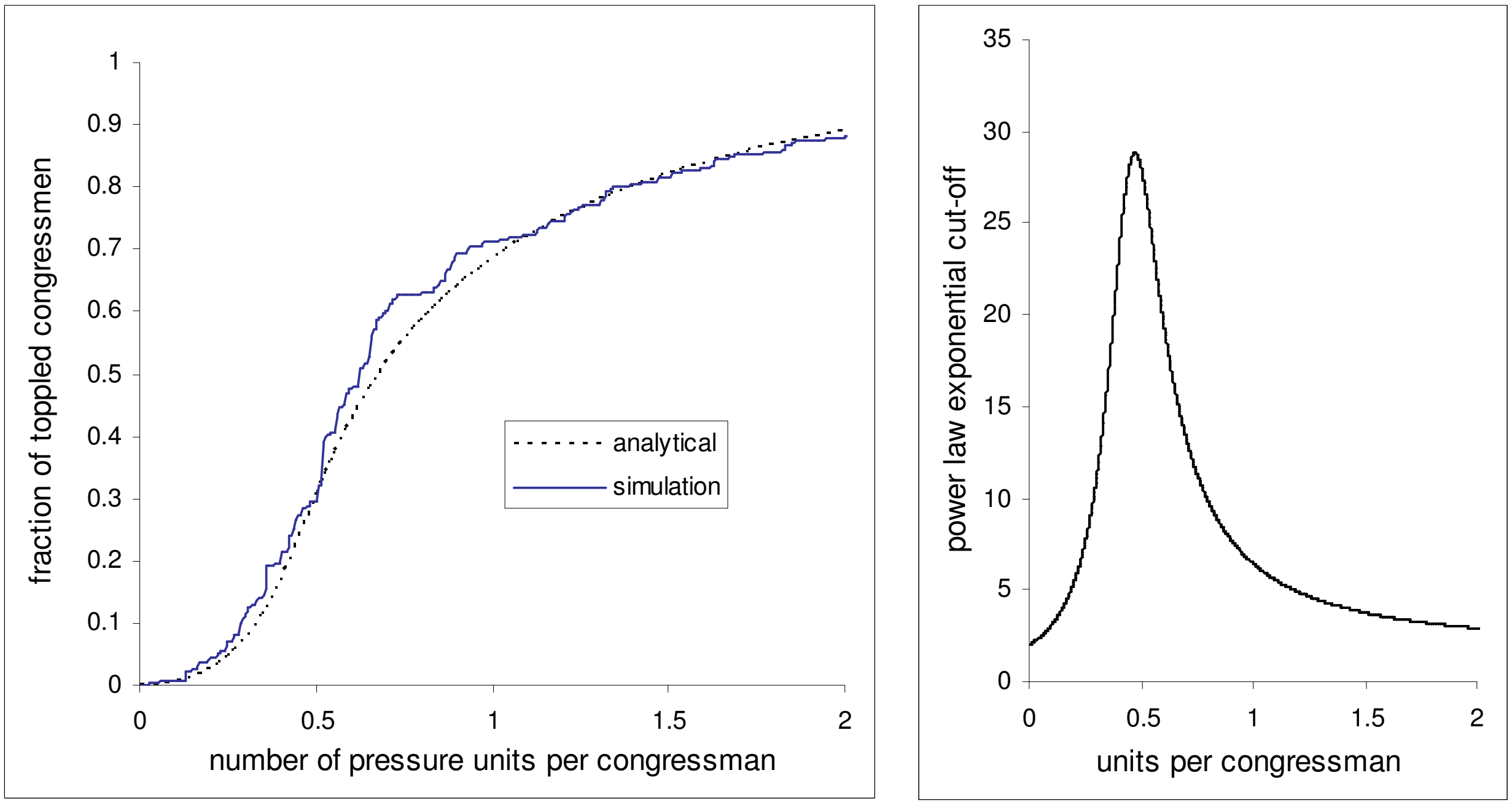

(a) (b)

**Figure 4.** (a) Fraction of toppled congressmen as a function of the number of units of political pressure per congressman. (b) Exponential cut-off of the power law of avalanche sizes.

The simulation curve follows the general tendency of the analytical curve, but in addition has a staircase structure. The analytical curve does not have steps because in an infinite system, there are always avalanches somewhere and the size of each avalanche becomes infinitely small, when divided by system's infinite size. A simulation of a hypothetical sandpile of a hundred thousand congressmen produced a curve almost indistinguishable from the analytical curve.

The distribution of the avalanche sizes depends on the parameter $r = 2p_1$. When it approaches 1, the system reaches a critical state where the average avalanche size diverges and the distribution of sizes becomes a power law. The numerical solution of Eq. (4) shows that $p_1$ reaches its maximum value of 0.37 at $m \approx 0.47$. This gives us $r_{\max} = 0.74$ which is not very close to the critical value $r_c = 1$. Why did we still get a power law in avalanche distribution? Let us take the formula for the probability distribution, $P(n)$, of avalanche sizes, $n$, in a random-network BTW model from Chapter 28 of Ref. [4]. After substituting $k = 2$ and $p(k) = p_1$ in Eq. (28.2) we get: $P(n) \propto \frac{1}{n^{3/2}} (4p_1(1-p_1))^{n/2}$. When $p_1$ is close to its critical value of ½, this equation has the asymptotic $P(n) \sim \frac{1}{n^{3/2}} \exp(-n/a)$, where the exponential cut-off of the power law is

$$a = 2/(1-2p_1)^2 . \quad (5)$$

For the maximum value $p_1 = 0.37$ we get $a = 29$. Fig. 4(b) shows $a$ as a function of $m$ computed using Eq.(5) and numerical solution of Eq. (4). For the period between $m \approx 0.34$ and $m \approx 0.66$ (this corresponds to two and a half month in Fig. 1) $a$ stays above 15. This is consistent with the maximum step height of 24 in the simulation and the maximum step height of 16 in the actual data (see Fig. 2(b)). A power-law fit ($P(n) \sim n^{-\alpha}$) to the step height distribution (Figure 3(b)) produces $\alpha \approx 1.1$ for actual data and $\alpha \approx 1.3$ for the simulation. When we restrict the data range to $n > 2$ we get $\alpha \approx 1.4$ for actual data and $\alpha \approx 1.9$ for the simulation. When there is not a lot of data, the estimated values of exponents are not very accurate and different ways of estimating exponents give different result. A different estimate obtained using numerical maximization of the likelihood function (Eq. (3.5) of Ref. [5]) gives the exponents $\alpha \approx 1.7$ for actual data and $\alpha \approx 1.8$ for simulation. This is consistent with the theoretical value $\alpha \approx 1.5$. The system for a sufficient period is close enough to its critical state to produce power-law distributed avalanches. In contrast, power law distribution of step length is not a property of the BTW model in a critical state. The longest steps occurred either at the beginning when $p_1$ was close to zero because most congressmen were in 0-unit state or after most congressmen toppled and $p_1$ became small again. We can derive the long step length asymptotic from our model. At small *m*, $p_1(m) \sim m$. Consequently the expected step length for given *m* is $l(m) = 1/p_1(m) \sim 1/m$. The probability that a randomly selected step corresponds to given *m* is $p(m) \sim 1/l(m) = m$. Now $p(l)$ can be obtained as

$$p(l) = p(m)\frac{dm}{dl} \sim \frac{1}{l^3}. \qquad (6)$$

This agrees with the simulation data shown in Fig.2(a), where the fit in the data range $l > 4$ gives the exponent of 2.7. The simulation of a hypothetical sandpile of 100,000 congressmen produced an even closer exponent of 3.1. The agreement of the actual experimental data with the theory is worse. The fit in the data range $l > 4$ gives the exponent of 2.3. This means that the theory, perhaps, correctly describes the functional form of the step length distribution, but gives a wrong value of the exponent.

Interestingly, we did not need to introduce any variance in the degree of influence of different congressmen to explain the data. In our model all congressmen have the same degree of influence. One might have suspected that the biggest steps of the staircase are due to joining of a highly influential congressman bringing with himself many new co-sponsors which he had influenced. In our model, big steps are a result of evolution of Congress to a sort of critical state, where any congressman can trigger an avalanche of co-sponsors. In a related study [6] major properties of scientific citation distribution could be explained assuming that all papers are created equal.

Note, however, that not all resolutions introduced in Congress get the same level of support. The difference between resolutions enters our model through public pressure parameter $\lambda$. If it is close to zero, the resolution will not get sizable support in reasonable time. An obvious extension to the model is to introduce political pressure against the resolution. We can model it by dropping at rate $\lambda_{neg}$ negative units of pressure in addition to positive units, which we drop at rate $\lambda_{pos}$. When positive pressure far exceeds negative pressure this model will give the same results as the model we used with dropping rate set to $\lambda = \lambda_{pos} - \lambda_{neg}$. An interesting case to investigate will be $\lambda_{pos} \approx \lambda_{neg}$. It could explain the cases when a resolution quickly gains some support, which, however, never becomes overwhelming.

Another possible explanation of the Devil's staircase was proposed by one of the referees of this paper. It may happen because the displays of co-sponsorship are precluded or constrained during recesses, holidays, and weekends. When we look at the data, we see that only five out of 313 co-sponsorships were announced during recesses, while recesses together with weekends and holidays take about half of the time. We simulated the following model. The 313 decisions to co-sponsor were randomly uniformly distributed over the time period starting with the date when the resolution was announced and ending with the last date for which we had the data. If the decision was taken during a recess, it is only announced on the first business day after the recess. Figure 6 shows the results of the simulation. It does indeed produce a Devil's staircase, though a peculiar one: the height of the next step is directly proportional to the length of the previous step. The correlation coefficient between the height of the step and the number of non-working days directly preceding the date of the step is 0.96. This is at odds with the actual data where the similar coefficient is -0.076. In addition, the distribution of step height does not match well the real data: the maximum step height produced by the simulation is 66 while the actual value is 16. On the other hand, the simulation produces a step-length distribution, which matches better the actual data than the congressional sandpile model. Perhaps, a more realistic model which takes into account both the recess constraints and inter-influence between congressmen will ideally match the data. This may be the subject of further investigations.

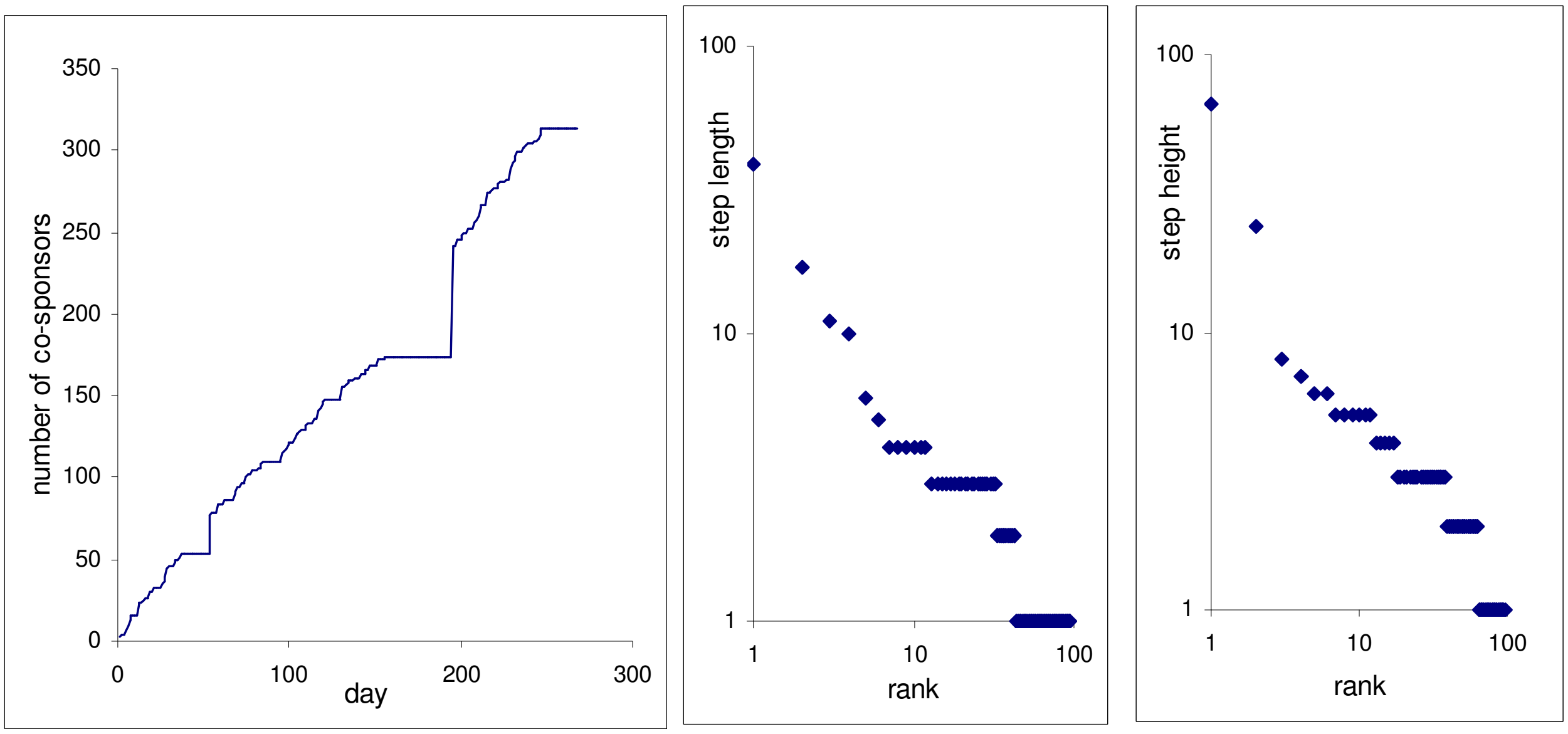

**Figure 5.** Numerical simulation of the model, which takes into account that displays of co-sponsorship is precluded during recesses/holidays/weekends but neglects any inter-influence between congressmen.